\documentclass{PoS}

\title{Three Neutrino Oscillations in Uniform Matter}

\ShortTitle{Three Neutrino Oscillations in Uniform Matter}

\author{\speaker{Ara Ioannisian}%\thanks{A footnote may follow.}
\\
        Yerevan Physics Institute, Alikhanian Br.\ 2, 375036 Yerevan,
Armenia\\
Institute for Theoretical Physics and Modeling, 375036
Yerevan, Armenia\\
        E-mail: \email{ara.ioannisyan@cern.ch}}

\author{Stefan Pokorski\\
        Institute of Theoretical Physics, Faculty of Physics, University of Warsaw, ul. Pasteura 5, PL-02-093 Warsaw, Poland\\
        E-mail: \email{Stefan.Pokorski@fuw.edu.pl }}

\abstract{Following similar approaches in the past, the Schrodinger equation for three neutrino propagation in
matter of constant density is solved analytically by two successive diagonalizations of 2x2 matrices.
The final result for the oscillation probabilities is obtained directly in the conventional parametric
form as in the vacuum but with explicit simple modification of two mixing angles ($\theta_{12}$ and $\theta_{13}$) and
mass eigenvalues.
}

\FullConference{ICHEP 2018, XXXIX International Conference on High Energy Physics\\
		4-11 July 2018\\
		COEX, Seoul, Korea}

\begin{document}

The MSW effect for  the neutrino propagation in matter attracts a lot of  experimental and theoretical attention. 

On the theoretical  side, a large number of numerical simulations of the MSW effect in matter with  a constant or varying density has been performed.  Although, in principle, sufficient for comparing the theory predictions with experimental data, they do not provide a transparent physical  interpretation
of the experimental results. Therefore, several authors have also published analytical or semi-analytical solutions  to the  Schroedinger equation for three neutrino propagation in matter of constant density, in various perturbative expansions ~\cite{ many papers, Blennow:2013rca,  Denton:2016wmg}.

We solve the Schroedinger equation in matter with constant density by two successive diagonalizations of 2x2 matrices (similar approaches have been used in the past, in particular in ref. \cite{Blennow:2013rca} and \cite{Denton:2016wmg}). The final result for the oscillation probabilities is obtained directly in  the conventional parametric form as in the vacuum but with modified  two mixing angles and mass eigenvalues\footnote {The results of this paper have been  presented as private communication by one of us (A.I) to the members of the T2HKK collaboration  in December 2017.}, similarly to the well  known results for the two-neutrino propagation in matter.
The three neutrino oscillation probabilities in matter have been presented in the same form  as here in the  recent ref.~\cite{PARKE ET AL}, where  the earlier results obtained in ref.
\cite{Denton:2016wmg}  are rewritten in this form. The form of our final results  can also  be obtained after some simplifications from ref.\cite{Blennow:2013rca}.
Our approach can be easily  generalized to  non-constant matter density by dividing the path of the neutrino trajectory in the matter to layers and assuming constant density in each layer.

 In the electroweak basis the neutrino Hamiltonian is
\begin{equation}
{\cal H}=U 
\left( \begin {array}{ccc}
 0&0&0\\ 
0&{\Delta m^2_\odot \over 2E}&0\\ 
0&0&{\Delta m^2_a \over 2E}
\end {array} \right)
U^\dagger +
\left( \begin {array}{ccc}
 V(x)&0&0\\ 
0&0&0\\ 
0&0&0
\end {array} \right) = 
U_m
\left( \begin {array}{ccc}
 0&0&0\\ 
0&{\Delta m^2_{21} \over 2E}&0\\ 
0&0&{\Delta m^2_{31} \over 2E}
\end {array} \right)
U^\dagger_m
%\nonumber
\label{electroweak}
\end{equation}
The matrix $U$ ($U_m$) is the neutrino mixing matrix in the vacuum (matter). The mass squared differences are defined as $\Delta m^2_\odot \equiv m^2_2-m^2_1$ ($\approx  7.5 \ 10^{-5} eV^2$) and 
$\Delta m^2_a \equiv m^2_3-m^2_1$ ($\approx  \pm 2.5 \ 10^{-3} eV^2$, positive sign is for normal mass ordering and negative sign for inverted one). ${\Delta m^2_{21} \over 2E}$ and ${\Delta m^2_{31} \over 2E}$ are eigenvalues of the neutrino Hamiltonian (we always can add proportional to unity diagonal matrix to the neutrino Hamiltonian).
$V(x)$ is the neutrino weak  interaction potential energy $V=\sqrt{2} G_F N_e$  ($N_e$ is electron number density) and we take it in this section to be x-independent.   

We work in the auxiliary basis \cite{Krastev:1988yu, Peres:2003wd, Ioannisian:2018qwl} and do two rotations for diagonalization of the neutrino Hamiltonian in eq \ref{electroweak}.

For the mixing angles $\theta_{13}^m$ and $\theta_{12}^m$ in matter we get 
\begin{equation}
\sin 2 \theta_{13}^m = {\sin 2 \theta_{13} \over \sqrt{(\cos 2\theta_{13}-\epsilon_a)^2+\sin^2 2 \theta_{13}}} , 
\hspace{1.2cm}
\cos 2 \theta_{13}^m \ = \ {\cos 2\theta_{13}-\epsilon_a \over \sqrt{(\cos 2\theta_{13}-\epsilon_a)^2+\sin^2 2 \theta_{13}}}, 
\label{t13}
\end{equation}
\begin{equation}
\sin 2 \theta_{12}^m = {\cos \theta^\prime_{13} \sin 2 \theta_{12} \over  \sqrt{(\cos 2 \theta_{12}-\epsilon_\odot)^2+\cos^2 \theta^{\prime}_{13} \sin^2 2 \theta_{12}} }, 
\ \ \
\cos 2 \theta_{12}^m  \ = \  { \cos 2 \theta_{12}-\epsilon_\odot \over  \sqrt{(\cos 2 \theta_{12}-\epsilon_\odot)^2+\cos^2 \theta^{\prime}_{13} \sin^2 2 \theta_{12}} }
\label{t12}
\end{equation}
where
\begin{equation}
\theta_{13}^\prime=\theta_{13}^m- \theta_{13}, \ \ \ \ \epsilon_a ={2 E V \over \Delta m_{ee}^2}, \ \ \ \ \epsilon_\odot ={2EV \over \Delta m^2_\odot}(\cos^2 \theta_{13}^m +{\sin^2 \theta^\prime_{13}\over \epsilon_a}), 
\ \ \ \ \Delta m^2_{ee}=  \Delta m^2_{a}- s_{12}^2 \Delta m^2_\odot
\end{equation}
And for differences between eigenvalues of the neutrino Hamiltonian,  $\cal H$,  we have
\begin{eqnarray}
\label{m21} 
{\Delta m^2_{21} \over 2 E}&=& {\Delta m_\odot^2 \over 2E} \sqrt{(\cos 2 \theta_{12}-\epsilon_\odot)^2+\cos^2 \theta^\prime_{13} \sin^2 2 \theta_{12}},
\\
\label{m31a}
{\Delta m^2_{31} \over 2 E}&=&
{3 \over 4}{\Delta m^2_{ee} \over 2E} \sqrt{(\cos 2\theta_{13}-\epsilon_a)^2+\sin^2 2 \theta_{13}} + \ {1 \over 4}[{\Delta m^2_{ee} \over 2 E} +V]
\ + \ {1 \over 4 E}(\Delta m^2_{21}-\Delta m^2_{\odot}\cos 2 \theta_{12})
\label{m31b}
%\nonumber
\end{eqnarray}
In our approximation 23 angle and the CP phase remain unchanged: 
$\theta_{23}^m \equiv \theta_{23}$ and $\delta^m \equiv \delta$ \cite{Petcov:2018zka}.

The oscillation probabilities $P_{\nu_\alpha \to \nu_\beta}$  ($\alpha, \beta = e, \mu, \tau$)  have the same forms as for the  vacuum oscillations with mass eigenstates as  above and with replacements $\theta_{12} \to \theta_{12} ^m $ and  $\theta_{13} \to \theta_{13} ^m $. For the $\nu_\mu\rightarrow \nu_e$ transition we have
\begin{eqnarray}
P_{\nu_\mu \to \nu_e} &=& \sin^2 2 \theta^m_{13} s_{23}^2 \  \left[ c_{12}^{m2} \sin^2 {\phi_{31} \over 2}+ s_{12}^{m2} \sin^2 {\phi_{32} \over 2}\right]
\nonumber
\\
&&
+{1 \over 2}c_{13}^m \sin 2 \theta^m_{13} \sin 2 \theta^m_{12} \sin 2 \theta_{23} \cos \delta \ \sin {\phi_{21} \over 2} \sin {\phi_{31} + \phi_{32}  \over 2} 
\nonumber
\\
&&
- c_{13}^m \sin 2 \theta^m_{13} \sin 2 \theta^m_{12} \sin 2 \theta_{23} \sin \delta \ \sin {\phi_{21} \over 2} \sin {\phi_{31}  \over 2} \sin {\phi_{32}   \over 2} 
\nonumber\\
&&
+\left[  c_{13}^{m2} \sin^2 2 \theta^m_{12} (c_{23}^2-s_{23}^2 s_{13}^{m2}) +{1 \over 4}c^m_{13} \sin 2\theta^m_{13}\sin 4\theta^m_{12}\sin 2\theta_{23} \cos \delta
 \right] \sin^2 {\phi_{21} \over 2}
 \label{Pmue}
\end{eqnarray}

This  approximate solution is valid for all energies. 
For  anti-neutrino oscillations $P_{{\bar \nu}_\alpha \to {\bar \nu}_\beta}$,  V$\to$ -V and $\delta \to -\delta$. 
For normal mass hierarchy $\Delta m_a^2$ is positive and for inverted mass hierarchy it is negative.

\vspace{-.35cm}
\begin{figure}[h]
\includegraphics[width=1\textwidth, height=0.4\textwidth]{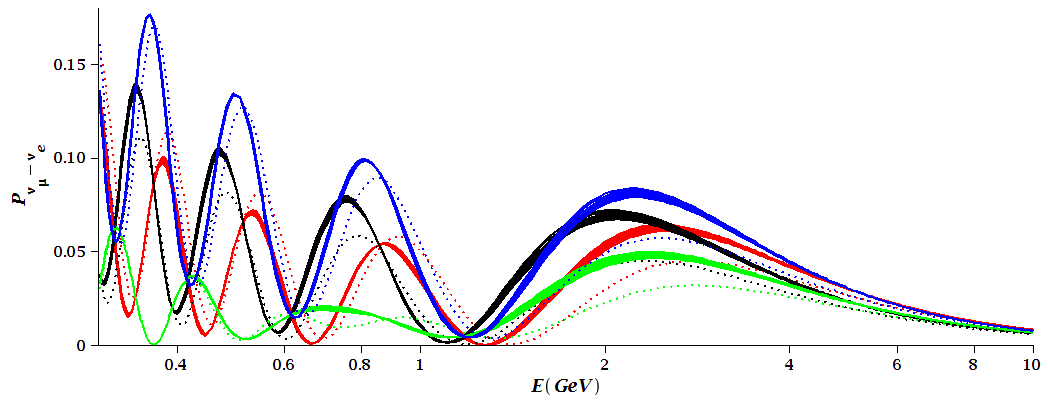}
\caption{$\nu_\mu \to \nu_e$ oscillation probability at DUNE for normal mass hierarchy, $\delta_{cp} = 0$ (red), $\delta_{cp} = {\pi \over 2}$ (green),  $\delta_{cp} = {\pi }$ (black), $\delta_{cp} = -{\pi \over 2}$ (blue). Thickness of the plots are from varying constant/uniform matter density 2.5 - 3 g/cm$^3$. Dotted plots are for  vacuum oscillations}
\label{NNH}. 
\end{figure}

Our solutions are illustrated in Fig.~\ref{NNH} 
%and~\ref{NIH} 
for $\nu_\mu \to \nu_e$ oscillation at DUNE distance for several  values of $\delta_{CP}$ and compared with the oscillation probabilities in the vacuum, shown by the dotted curves.

The most important effect is the dependence of the oscillation  probability on  the angle $\theta_{13}$
which  has larger (smaller) values in matter than in the vacuum for normal (inverted)  neutrino mass hierarchies (and opposite for antineutrinos). Thus the oscillation probabilities have larger(lower) oscillation amplitudes for normal (inverted)  neutrino mass hierarchies (and opposite for antineutrinos). In oder words the matter of the Earth is amplifying the effect of the mass ordering on neutrino oscillations. The dependence on the angle $\theta_{13}$ enters multiplicatively in the first three terms of eq (\ref{Pmue}), whereas the fourth term is small in the region of the first maximum.
Therefore the matter effects relative to the oscillations in the vacuum do not depend on the value of $\delta_{CP}$, as it is seen in Fig.~\ref{NNH}
 %and~\ref{NIH}
. Moving to the next resonances (lower energies) the difference between
oscillations in matter and in the vaccuum remain qualitatively similar, although   some small differences can be seen due to the fact that the change in the angle $\theta_{13}$ is smaller.

\begin{figure}[h]
\includegraphics[width=1\textwidth, height=0.4\textwidth]{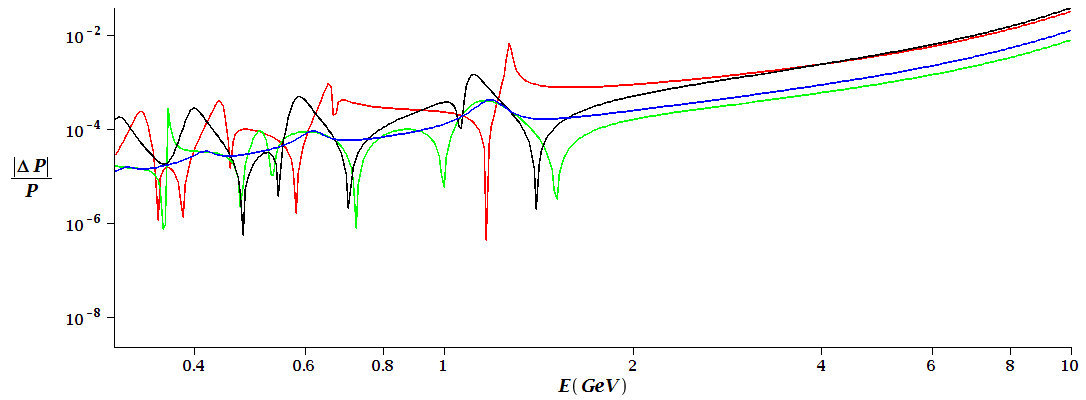}
\caption{${|\Delta P| \over P}  \equiv {|P^{num}_{\nu_\mu \to \nu_e}-P^{anl}_{\nu_\mu \to \nu_e}| \over P^{num}_{\nu_\mu \to \nu_e}}$. The relative error of our analytic result to the exact (numeric) $\nu_\mu \to \nu_e$ oscillation probability for normal mass hierarchy, $\delta_{cp} = 0$ (red), $\delta_{cp} = {\pi \over 2}$ (green), $\delta_{cp} = {\pi }$ (black), $\delta_{cp} = -{\pi \over 2}$ (blue). Matter density 2.6 g/cm$^3$. }
\label{RENN}. 
\end{figure}

In Fig.~\ref{RENN} we show the accuracy of the analytical solutions  comparing them with numerical/exact results.

\vspace{-0.35cm}
%\begin{figure}[h!]
%\includegraphics[width=1\textwidth, height=0.4\textwidth]{NIH.png}
%\caption{$\nu_\mu \to \nu_e$ oscillation probability at DUNE for inverted mass hierarchy, $\delta_{cp} = 0$ (red), $\delta_{cp} = {\pi \over 2}$ (green), $\delta_{cp} = {\pi }$ (black), $\delta_{cp} = -{\pi \over 2}$ (blue). Thickness of the plots are from varying constant/uniform matter density 2.5 - 3 g/cm$^3$. Dotted plots are for  vacuum oscillations}
%\label{NIH}. 
%\end{figure}

\vspace{-1.2cm}
%{\bf Acknowledgements.} It is pleasure to thanks the organizers of the ICHEP2018 for organizing a very interesting and enjoyable conference.

\end{document}